\documentclass[modern]{aastex631}

\usepackage{amsmath}

\newcommand\uf{Bryant Space Science Center, Department of Astronomy, University of Florida, Gainesville, FL 32611, USA}
\newcommand{\tess}{{\it TESS}}

\newcommand{\Kepler}{{\it Kepler}}

\newcommand{\rearth}{{\ensuremath{R_{\oplus}}}}
\newcommand{\aong}{{\ensuremath{\alpha/|g|}}}

\newcommand{\msun}{{\ensuremath{M_{\odot}}}}

\newcommand{\teff}{{\ensuremath{T_{\rm eff}}}}
\newcommand{\teq}{{\ensuremath{T_{\rm eq}}}}

\submitjournal{ApJ}

\begin{document}

\title{Plausibility of Capture into High-Obliquity States for Exoplanets in the M Dwarf Habitable Zone}

\correspondingauthor{Natalia Guerrero}
\email{natalia.guerrero@ufl.edu}

\author[0000-0002-5169-9427]{Natalia M.~Guerrero}
\affiliation{\uf}

\author[0000-0002-3247-5081]{Sarah A.~Ballard}
\affiliation{\uf}

\author[0000-0001-8283-3425]{Yubo Su}
\affiliation{Department of Astrophysical Sciences, Peyton Hall, 4 Ivy Lane, Princeton, NJ 08544, USA}


\begin{abstract}
For temperate exoplanets orbiting M dwarf hosts, the proximity of the habitable zone to the star necessitates careful consideration of tidal effects. Spin synchronization of the planetary orbital period and rotation period, tidal locking, and the subsequent impact on surface conditions, frames common assumptions about M dwarf planets. We investigate the plausibility of capture into Cassini State 2 (CS2) for a known sample of 280 multiplanet systems orbiting M dwarf hosts. This resonance of the spin precession and orbital precession frequencies can excite planets into stable nonzero rotational obliquities, breaking tidal locking and inducing a version of ``day" and ``night’’.  Considering each planetary pair and estimating the spin and orbital precession frequencies, we find 75\% of detected planets orbiting M dwarfs may be plausibly excited to a high obliquity and maintain it through subsequent tidal dissipation over long timescales. We also investigate two possible mechanisms for capture into CS2: quantifying the orbital migration or primordial obliquity necessary for CS2. We find orbital migrations by a factor of $\lesssim$2 and an isotropic initial spin distribution can produce high-obliquity planets, aligning with similar findings for planets orbiting close-in to FGK dwarfs. Many of the planets in our sample reside in both CS2 and within their stellar habitable zone. Over half of planets with $\teq<400$ K around host stars with $\teff<3000$ K could possess non-zero obliquity due to residence in CS2. This overlap renders the potential capture into Cassini States extremely relevant to understanding the galaxy's most common temperate planets.
\end{abstract}



\section{Introduction} \label{sec:intro}

The abundance of M dwarf planets in the Milky Way makes them compelling sites for habitability investigations. Planets smaller than Neptune are are 3.5 times more abundant around M dwarfs as compared to Sunlike stars \citep{Mulders15, Dressing15}, and small stars themselves comprise $75\%$ of stars in the Galaxy \citep{Henry04}.  The fact that smaller and cooler planets orbiting M dwarfs present much larger signals in comparison with those orbiting Sun-like stars \citep{Tarter07}, also makes them attractive targets for follow-up efforts. However, the environment for planets around M dwarf stars is very different from that for Sun-like stars:  among many relevant effects, the planets receive a very different budget of photons from the host star (see \citealp{Shields16} for a summary comparison). The lower luminosity of M dwarfs also necessitates a ``habitable zone" much closer to the star. For example, a planet orbiting a $0.25 \msun$ star must orbit only 0.1 AU away to receive Earth-like energy insolation \citep{Kopparapu13}. 

It is this proximity of potentially habitable planets to the host M dwarf that makes dynamical information especially important. Tides, whether due to planetary orbital eccentricity or spin obliquity, play an outsize role for potentially temperate planets orbiting M dwarfs. On Earth, tidal dissipation contributes heat at a level one-millionth that of the incident energy from the Sun \citep{Luna21_Earthtides}. But on a planet in the habitable zone of the $0.25 \msun$ star, tides from obliquity alone can contribute $1\%$ of the heat budget of the planet \citep{Millholland19_obliquity}, and the tidal heating from an orbital eccentricity of $\sim0.1$ deposits sufficient heat inside the planet to induce a runaway ``tidal Venus'' effect \citep{Barnes13}. 

Even establishing whether an M dwarf planet resides in the ``habitable zone'' of its star relies crucially on dynamical information \citep{Barnes08, Yang14}. In conceptualizing M dwarf planetary systems, the model of Jupiter and the Galilean moons is useful, firstly in terms of relative scale (the Kepler-961 system of a 0.13 $M_{\odot}$ star and three terrestrial planets fits inside the same footprint, per \citealt{Muirhead12a}). Secondly, the dynamical factors that conspire to make Io the most volcanically-active body in the Solar System are also relevant in M dwarf planetary systems: even the very low eccentricity of 0.0041, forced by resonant interactions between Io, Ganymede, and Callisto, produces enormous tides on Io that resurface it entirely every $\sim$1 million years \citep{peale1979b, hamilton2013}. 


Among many important implications for the strength of tidal interactions on M dwarf habitable zone planets, spin synchronization is commonly considered a foregone conclusion \citep{Yang14}. The potential threat to habitability of a permanent cold night side and a roasting day side could be severe: even when the permanent night side doesn't threaten outright atmospheric collapse \citep{Heng12}, it can be an efficient cold-trap for water \citep{Menou13} and even strongly destabilize the carbonate-silicate cycle \citep{Edson12}. 

However, tidal locking need not occur if the planet possesses a tilted, or oblique, axis of rotation. 
A hypothetical oblique planet would not be tidally locked but would rotate subsynchronously, with a sidereal day longer than its year: such a scenario would have significant implications for the atmospheric heating and circulation on the planet \citep{shan2018, quarles2020, Saillenfest2019}. 
However, the \textit{a priori} likelihood for this scenario, considering a single planet and star in isolation, is low. Tidal energy dissipation should drive the planet's obliquity and eccentricity to zero over timescales of $10^7$-$10^8$ years \citep{Heller11_tides, Wisdom08, Millholland19_radius}. To maintain an obliquity over long timescales would therefore require an external driving force \citep{Fabrycky07a}. 
Alternately, planets which form with a near-zero obliquity may be driven to high obliquity by chaotic interactions, as may have occurred over the lifetime of the Solar system for the inner planets \citep{laskar1993, touma1993}.
Previous studies have considered pairs of M dwarf planets near mean-motion resonance which can cause migration of the substellar point, though these studies assume the planets have an obliquity value of zero \citep{Vinson17}. 

\citet{Millholland19_obliquity} demonstrated that the driver behind a long-term high obliquity could be spin-orbit resonance between planetary pairs. Specifically, capture into a specific Cassini State (involving synchronous precession of the planetary spin and orbital angular momentum vector) could allow planets to maintain non-zero obliquities for long timescales, while paradoxically enhancing tidal heating by a factor of $10^{3}$-$10^{4}$ \citep{Peale1969, Wisdom08, Su2022a}. 

A ``Cassini State" is a spin equilibrium resulting from the competition between the tidal torque (driving spin-orbit alignment and synchronization) and the gravitational torque from a neighbor (which drives orbital precession of the planet; \cite{Peale1969, Su2022a}). The Solar System provides multiple case studies of Cassini States, with the Moon's motion being the first explored with this framework \citep{Kaula68}. The generalization to other Solar System bodies followed soon after (\cite{Colombo66}; including the use of the term ``Cassini State" per \citealt{Peale1969}). The second Cassini State (hereafter CS2) is one of either two or four possible configurations, with both the possibility and the stability of the configuration contingent on the ratio between $\alpha$, the  spin-axis precession frequency, and $g$, the nodal precession frequency \citep{Ward1975a, Su2022a}. The evolution of the two relevant frequencies to commensurability determines which Cassini State results: of the resulting possible spin equilibria, CS2 allows for the steady maintenance of high obliquity over long timescales \citep{Correia15, Millholland19_obliquity, Su2022a}. The compact nature of multi-planet systems of super-Earths, particularly common among M dwarfs \citep{Muirhead15, Ballard16}, may make such commensurabilities more common: \citet{Su2022b} posited that the ubiquity of compact systems may lead to significant planetary obliquities as the norm. 

It's therefore possible, via the mechanism of capture into CS2, that some M dwarf planets may have high forced obliquities and resist the 1:1 tidal locking effect for billions of years. These planets would still experience some version of day and night, which combined with non-zero obliquity, would result in complicated surface insolation and atmospheric circulation \citep{Dobrovolskis13, ohno2019a}. Typically, capture into CS2 is physically plausible only for close-in planets, where the spin and orbital precession frequencies of planets are likelier to overlap (see e.g. \citealt{Su2022a, Millholland19_obliquity}). It is the proximity of the habitable zone to an M dwarf host star that makes this effect potentially relevant to temperate planets, as with other tidal effects. In addition to the implications for day/night rotation, the existence of seasonal cycles is also thought to enhance nutrient recycling, biospheric productivity, and biosignature detectability \citep{Barnett22, Jernigan23}. Small, closely-packed planets around M dwarfs are abundant in the Galaxy and the possibility of high obliquity could significantly change the conditions for habitability on close-in M dwarf planet. 

We aim here to investigate the plausibility of this effect among M dwarf planetary systems with a comparison between the likely spin and orbital precession frequencies for planets in known M dwarf multi-planetary systems. Due to the high degree of uncertainty for many of the planetary physical properties, we consider the problem at an order-of-magnitude scale. We aim here to estimate the relevant frequencies, consider their commensurability, and share resulting predictions for the implied planetary obliquity and rotation rates. Though our approach is simplified, we do consider some additional details in our investigation of the plausibility of CS2 for this sample of planets. Namely, we consider the resulting stability of the CS2 state, given that planets excited to the state may subsequently lose the high obliquity via tidal dissipation. We also comment upon the feasibility of two possible scenarios leading to the capture into CS2: capture during migration, and capture from a high primordial obliquity. 

This manuscript is organized as follows. In Section \ref{sec:methods}, we describe our sample of multiplanet systems. We summarize the framework that we employ to estimate (1) the individual planetary spin precession frequencies and (2) the nodal precession frequencies for each set of neighboring pairs. We describe how we then infer the plausibility of excited obliquity, and by extension, the associated planetary spin rate. In Section \ref{sec:analysis}, we consider the estimated frequencies at a population level and the potential for frequency commensurability. We examine the specific predictions for the TRAPPIST-1 system. We explore the implied spin rotation rates attributable to the estimated obliquities for pairs of planets for which capture into CS2 is plausible. In this Section, we also consider the stability of the CS2 state given tidal dissipation resulting from the high obliquity. We also consider two physical scenarios for capture into CS2, and investigate the conditions required in each scenario for excitation to occur. Finally, by comparing the range of orbital separations in the nominal ``habitable" zone to those with the highest likelihood of obliquity excitement, we identify the stellar spectral type where this overlap is strongest. In Section \ref{sec:conclusions}, we discuss the relevance of our findings for the population of M dwarf exoplanets and their habitability, and next steps for this study.




\section{Methods}
\label{sec:methods}

To test the feasibility of capture into CS2 among known M dwarf planetary systems, we employ the secular framework set forth by \cite{Millholland19_obliquity}. In Section \ref{sec:sample}, we first describe our sample of known planetary systems. In order for excitation to CS2 to occur, multiple conditions must align: some conditions are determined by a planet's individual properties, and others by the planet's separation from its neighbors. The plausibility of this hypothetical spin-orbit resonance depends upon our estimation of two frequencies. We describe these frequencies and our techniques for estimating them in Section \ref{sec:estimation}. An intriguing implication of high obliquity is the resistance to 1:1 tidal locking. The equilibrium rotation rate, if capture into CS2 has occurred, can be determined from the combination of obliquity and the planet's orbital frequency. We set out our framework for translating obliquity into rotation rate in Section \ref{sec:rotation_from_obliquity}. 

\subsection{Sample Selection}
\label{sec:sample}

\begin{figure*}[ht!]
    \centering
    \includegraphics[width=\textwidth]{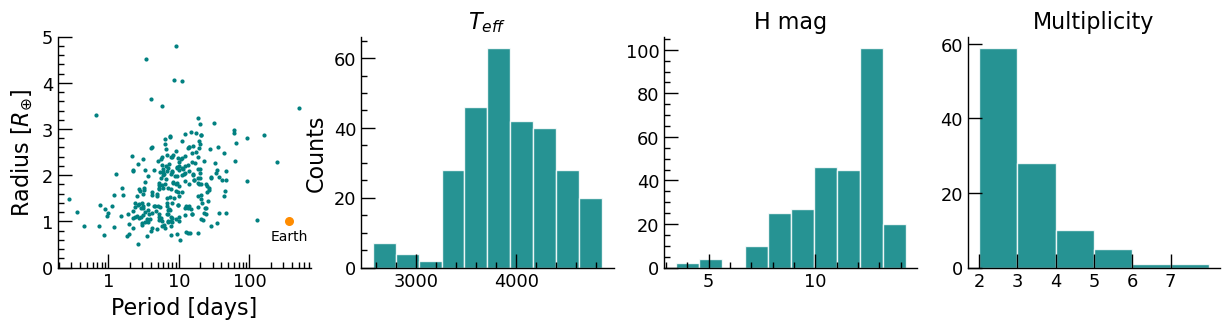}
    \caption{Sample of confirmed exoplanets orbiting M dwarf host stars. Our sample consists of planets with $R_p = 0.5-16.8 \rearth$, with the majority smaller than Neptune and with periods less than 300 days (first panel). The stellar temperatures (second panel) vary between 2560 and 4900 K. The H-band magnitudes (third panel) vary between $14.2$ and $3.47$. Lastly, we only consider multi-planet systems (fourth panel) with between two and seven planets in the system.
    }
    \label{fig:sample}
\end{figure*}

We chose our sample from the Exoplanet Archive catalog of confirmed exoplanets. 
As illustrated in Figure \ref{fig:sample}, we chose a sample of specifically small planets in multi-planet systems around cool stars ($\teff < 4700K $)\footnote{NASA Exoplanet Archive, \doi{10.26133/NEA12}, accessed June 2022}. We define ``multi-planet" to mean more than one detected transiting planet orbiting the star. 
Our sample contains 104 M dwarf hosts to a total of 280 planets. Among these hosts are 59 two-planet systems, 28 three-planet systems, 10 four-planet systems, 5 five-planet systems, and 1 each of six- and seven-planet systems. The long observational baseline and high photometric precision furnished by NASA's \Kepler~ Mission \citep{Borucki11, Gilliland11} enabled a high yield of multi-planet systems \citep{Fabrycky12b} and furnishes half of our sample: 143 planets in 51 systems. In addition, we include 137 planets in 53 systems discovered by \textit{K2}, \tess, and multiple ground-based observatories. Figure \ref{fig:sample} depicts the properties of the planetary sample (in radius versus orbital period, as well as multiplicity) and of the stellar sample (effective temperature, brightness). 

To calculate the frequencies relevant for potential capture into CS2, we rely on a combination of (1) the measured planetary and orbital parameters for the systems in our sample and (2) a suite of assumptions for unknown quantities (we detail this process below). Orbital period and planet radius, in addition to stellar mass and radius, are observationally constrained by the transit lightcurves and spectroscopy of the host stars: the average fractional uncertainty for $R_p/R_{*} = 0.068$ and the average fractional uncertainty for planet orbital period is $ 4.47 \times 10^{-5}$. The masses and radii of the host stars in our sample, inferred either from near-infrared or optical spectroscopy, have typical uncertainty of 10\% \citep{Muirhead12a, Swift15, Mann13}.





\subsection{Estimation of Relevant Frequencies for Sample}
\label{sec:estimation}
We calculate $\alpha$, the  hypothesized spin-axis precession frequency, and $g$, the nodal precession frequency, using the expressions for $\alpha$ and $g$ laid out in \cite{Millholland19_obliquity}. Frequency $\alpha$, expressed in Equation \ref{eq:alpha_spin}, is determined on a per-planet level and depends upon $M_*$, the stellar mass, $M_p$, 
the planetary mass, $R_p$, the planet radius, $a$, the semi-major axis, $k_2$, the Love number, C, the moment of inertia normalized by $m_{\rm p}R_{\rm p}^2$, and $\omega$, the spin frequency, where $\omega = 2\pi/P$ for $P$, the orbital period.

\begin{equation}
    \alpha = \frac{1}{2}\frac{M_*}{M_p}\left(\frac{R_p}{a}\right)^{3}\frac{k_2}{C}\omega   
    \label{eq:alpha_spin}
\end{equation} 

 We note here that we employ the simplifying estimation of setting $\omega=2\pi/P=n$ for this first-order estimation of $\alpha$ as in \cite{Millholland19_obliquity}; while we go on to explore resulting $\omega$ values other than $n$ eventually, this $\omega$ is useful as a starting benchmark value. Similarly, we employ the \cite{Millholland19_obliquity} estimation for the Love number, $k_2$. Because it is unknown, we draw a random value from a uniform distribution, $k_2 \in [0.1, 0.6]$. We compute C using the Darwin-Radau approximation given in Equation \ref{eq:C} \citep{Kramm11}. 

\begin{equation}
    C = \frac{2}{3}\left[1 - \frac{2}{5}\left(\frac{5}{k_{2} + 1} - 1 \right)^{1/2} \right]
    \label{eq:C}
\end{equation}

For planets not residing in mean-motion resonance (MMR), the estimated orbital precession frequency $g$ is set to the Laplace-Lagrange frequency $g_{\rm LL}$ expressed in Equation \ref{eq:g} (that is, $g/g_{\rm LL} = 1$ unless planets are near MMR). We refer specifically here to the 3:2 and 2:1 MMRs, which we define as $P_2/P_1 \in [1.5,1.55]$ or $[2.0, 2.05]$. To calculate $g$ for planetary systems which are near MMR, we modify the Laplace-Lagrange frequency per \cite{Millholland19_obliquity}, scaling $g_{\rm LL}$ by a randomly drawn constant so that the eventual $g/g_{\rm LL} = [0.3-1.0]$. 

\begin{equation}
  g_{\rm LL} = -\frac{1}{4}b_{3/2}^{(1)}(\alpha_{12})\alpha_{12} \left(n_{1}\frac{m_{2}}{M_* + m_1}\alpha_{12} + n_2\frac{m_1}{M_* + m_2}\right)
  \label{eq:g}
\end{equation}

We note that this $\alpha$ is distinct from Equation \ref{eq:alpha_spin}. Rather, $\alpha_{12}$ is the ratio of orbital separations of the two planets $a_1 / a_2 $. The mean-motion of the $i$th planet $n_i=2\pi/P$, and $b_{3/2}^{(1)}(\alpha_{12})$ is a Laplace coefficient given by Equation \ref{eq:b} (in practice, we compute $b_{3/2}$ using \texttt{pylaplace.LaplaceCoefficient}).

\begin{equation}
    b_{3/2}^{(1)}(\alpha_{12}) = \frac{1}{\pi} \int_{0}^{2\pi} \frac{\cos \psi}{(1 - 2\alpha \cos \psi + \alpha^2)^{3/2}}d\psi
    \label{eq:b}
\end{equation}

\subsection{Inference of Rotation Rate from Obliquity}
\label{sec:rotation_from_obliquity}

\begin{figure}[ht!]
    \centering
    \includegraphics[width=3in]{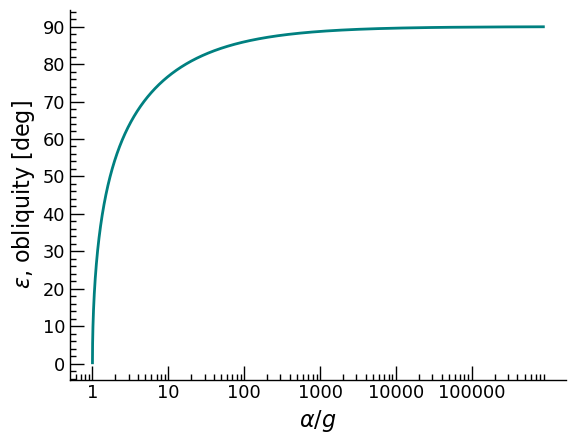}
    \caption{Obliquity rapidly goes to zero when nodal recession frequency, $g$, is greater than orbital precession constant, $\alpha$. The horizontal axis shows calculated values of \aong~for our sample, with the corresponding obliquity, $\epsilon$, in degrees, on the vertical axis, calculated using Equation \ref{eqn:obliquity}.}
    \label{fig:ag_vs_obliquity}
\end{figure}

We go on to estimate the resulting obliquity of the planet, which can be calculated with the ratio \aong. Because we consider here orbital precession of circular, prograde orbits (that is, with mutual inclination $<$90$^{\circ}$), we assume $g$ is negative for the purposes of this study. We note that a relaxation of this assumption, such as considering retrograde orbits or eccentricity would allow for the possibility of positive g \citep{Barnes16, Kreyche20}. For the sake of exploring this hypothesis, we assume that the evolution toward this resulting \aong~has resulted in capture into the Cassini 2 state. In truth, the resultant Cassini State may not necessarily be State 2 depending on the details of the approach to commensurability, per \citealt{Su2022a}. For planets in Cassini states, the threshold obliquity beyond which tidal locking occurs is described by Equation \ref{eqn:obliquity} from \cite{Millholland19_obliquity} where $\rm{\alpha_{\rm syn}} = \alpha(n/\omega)$ for the case of synchronous rotation, $\omega = n$. Figure \ref{fig:ag_vs_obliquity} shows the calculated values for $\epsilon$ from \aong. Non-zero obliquities in this state are only possible for $\aong> 1$ where $\alpha > |g|$. 

\begin{equation}
    \cos \epsilon = \left(\frac{1}{2\rm{\alpha_{\rm syn}}/|g| - 1} \right)^{1/2}
    \label{eqn:obliquity}
\end{equation} 

In Figure \ref{fig:ag_vs_obliquity}, we represent the resulting calculated \aong~and $\epsilon$ values for our sample, per the functional form of Equation \ref{eqn:obliquity}. It is useful to benchmark the spin obliquity induced by \aong~$>$ 0 states: an \aong~of 2.5 corresponds to a spin obliquity in CS2 of $\sim 60^{\circ}$. For ratios of $\aong~\approx 1-100$, the range of obliquity values are $\epsilon \approx 0-90^{\circ}$, with obliquity asymptotically approaching $90^{\circ}$ for \aong~$>100$. 

\cite{Millholland19_obliquity} lays out a viscoelastic model for obliquity tides, from which it is possible to then calculate $\omega_{\textrm{eq}}$, the equilibrium rotation rate of the planet as a function of obliquity (where $d\omega/dt = 0$). This functional form is given by Equation \ref{eq:epsilon_to_w}:

 \begin{equation}
     \frac{\omega_{\textrm{eq}}}{n} = \frac{N(e)}{\Omega(e)}\frac{2\cos~\epsilon}{1 + \cos^2\epsilon}.
     \label{eq:epsilon_to_w}
 \end{equation} 

\begin{figure}[ht!]
    \centering
    \includegraphics[width=3in]{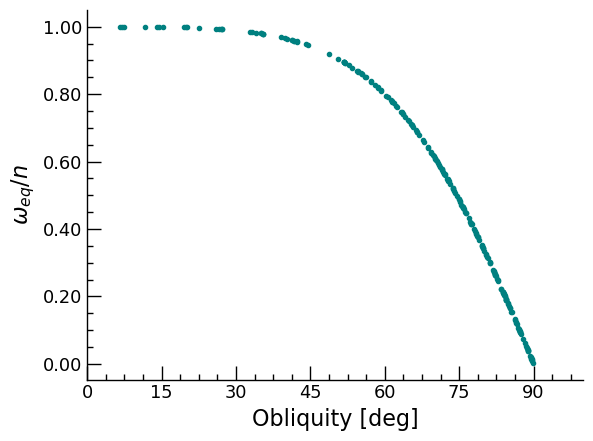}
    \caption{Oblique planets in CS2 rotate subsynchronously. The horizontal axis plots our predicted values for obliquity. The vertical axis shows the $\omega_{\textrm{eq}}/n$ values calculated using Equation \ref{eq:epsilon_to_w}.}
    \label{fig:weqn_obl}
\end{figure}

We assume e$\sim$0 in systems of densely populated transiting planets: a justifiable assumption per dynamical stability arguments \citep{Jontof15}, in addition to which \cite{Sagear23} showed that orbiting eccentricities in multi-planet systems orbiting M dwarfs possess typical $e=0.03^{+0.02}_{-0.01}$. In this special $e=0$ case, the functions $ N(e=0) = \Omega(e=0) = 1$. Figure \ref{fig:weqn_obl} shows the resulting equilibrium rotation rates for our sample, using the calculated $\epsilon$ values from Equation \ref{eqn:obliquity}. The ratio of $\omega_{\textrm{eq}}/n$ approaches 1, the value corresponding to tidal locking, as obliquity, $\epsilon = 0^\circ$. For higher values of obliquity, when coupled with $e=0$, planets experience subsynchronous rotation.

\section{Analysis}
\label{sec:analysis}

We now consider our findings as a whole, after applying the the methodology described in Section \ref{sec:methods} to our sample of known M dwarf multi-planet systems. In Section \ref{sec:frequencies}, we compare the two relevant frequencies: the estimated orbital precession frequency $g$ for each planet pair, and their spin precession frequencies $\alpha$. From these estimates, we quantify at a population level the fraction of cases in which these frequency distributions overlap. In Section \ref{sec:obliquities}, we go on to estimate the degree of obliquity excitation for subsample of the planets with favorable $\alpha/|g|$ ratios. Combining these obliquity values with their planetary orbital periods in Section \ref{sec:rotation}, we infer the resulting planetary rotation rates. 

We investigate the effects of tidal damping in Section \ref{ss:capture_cs2}: only a subset of planets excited to the CS2 state can stably remain there, depending on the tidal dissipation rate as compared to the characteristic tidal timescale on which the spin and rotation of the planet evolve. In Section \ref{ss:capture_cs2}, we consider in more detail the physical mechanisms can induce CS2. These include the inward migration of planets during formation, or alternatively capture into high-obliquity CS2 as a result of a high primordial obliquity. Having demonstrated the plausibility of capture into CS2 for a subset of planets in our sample, and their subsequent stability, we consider the context of potential habitability. In Section \ref{sec:cassini_hz}, we compare two relevant orbital separation ranges: (1) the orbital separations that are likeliest to result in CS2 capture and (2) the nominal range of orbital separations that bracket the ``habitable zone." We investigate the dependence of this overlap on stellar effective temperature. 

\subsection{Frequency Commensurability}
\label{sec:frequencies}
We determine the likelihood of the relevant resonance, between the spin precession frequency $\alpha$ and the nodal precession frequency $g$, for each of 274 known planetary pairs (that is, considering only pairs of nearest neighbors). To create these distributions, we randomly draw an $\alpha$ value from the range of values calculated for each planet using a set of randomly-generated values for the Love number, $k_2$ (described in Section \ref{sec:methods}). Each system in our sample hosts at least two transiting planets, allowing us to estimate the nodal precession rate, $g$, which is calculated for each pair of planet interactions. For systems with $N > 2$  planets, each system will furnish $N-1$ values for $g$ and $N$ values for $\alpha$ (the increased number of overlap opportunities makes multi-planet systems much likelier sites for CS2 capture, per \citealt{Su2022b}). 

\begin{figure}[ht!]
    \centering
    \includegraphics[width=3in]{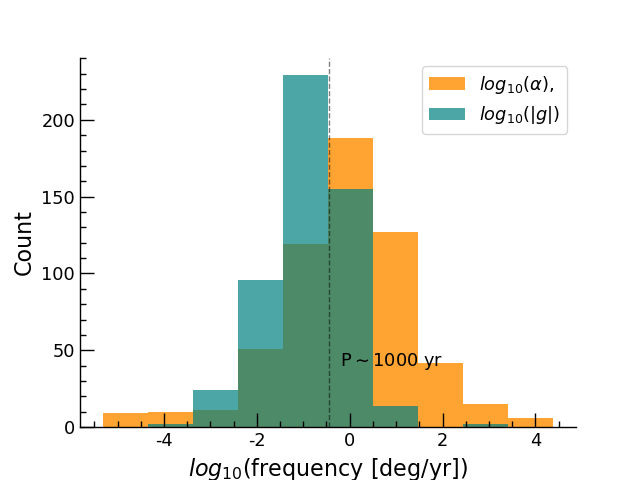}
    \caption{Roughly half of the values of $\alpha$ and $g$ for our planet sample overlap with a peak value roughly corresponding to a period of $\sim1000$ years. For each planet in the sample, we calculate $\alpha$, the spin-axis precession constant, and $g$, the nodal recession frequency. All values for $g$ are shown in the histogram.  Note the commensurability of the two distributions, making Cassini State capture plausible.}
    \label{fig:histogram_g_alpha}
\end{figure}

As shown in Figure~\ref{fig:histogram_g_alpha}, we find that the population-level distributions between $\alpha$ and $|g|$ overlap at the level of 50\% (note that the frequencies are shown in degrees $\times$ year$^{-1}$). The likeliest frequency for $\alpha$ corresponds to a precession period of roughly $1,000$ years, an order of magnitude shorter than the precession period for the spin axis of Earth, which is $\sim24,000$ years \citep{hays1976}. 

We identified a benchmark planetary system, Kepler-32 \citep{Swift13}, that is included in both \citet{Millholland19_obliquity} (whose methodology we employ here) and in our own sample. We established consistency in our  calculated values for $\alpha$ and $g_{\rm LL}$ for Kepler-32 b and c with the values computed for the same planet pair in \cite{Millholland19_obliquity}, when drawing planet masses $M_p$ from \citet{haddenlithwick2017}. We benchmarked against their findings $g_{\rm LL}$ for the pair ($1.13 \times 10^{-10} s^{-1}$) and $\alpha$ for each planet ($1.62 \times 10^{-8}$ and $ 4.99 \times 10^{-9} s^{-1}$, respectively). 
Following this benchmark confirmation, we calculated the planet masses necessary for Equations \ref{eq:alpha_spin} and \ref{eq:g} by employing the mass-radius relations of \citet{ChenKipping18} with our radii from the Exoplanet Archive Sample described in Section \ref{sec:sample}.

\subsection{Resulting Obliquities for Sample}
\label{sec:obliquities}

\begin{figure*}[ht!]
    \centering
    \includegraphics[width=5in]{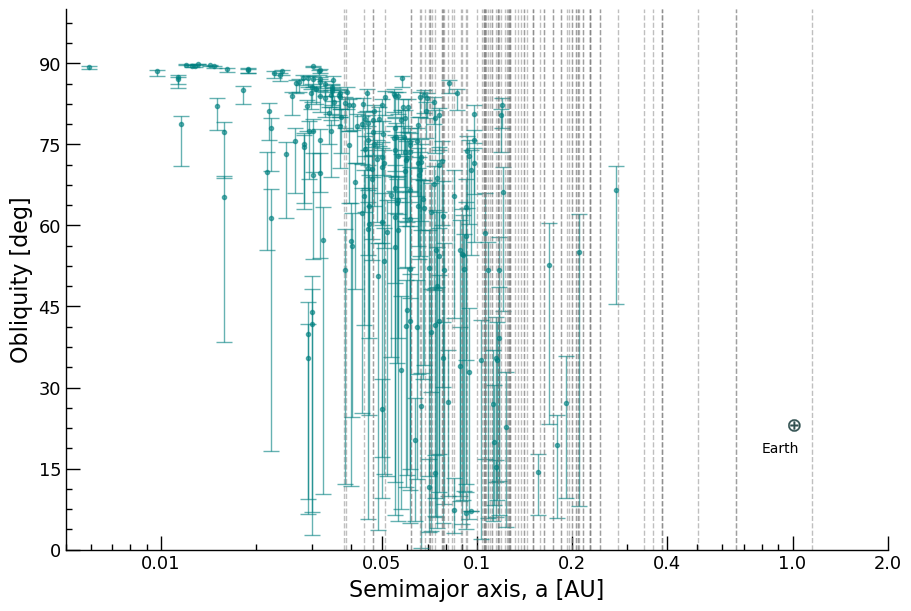}
    \caption{Calculated obliquity ranges for each planet, as a function of orbital semi-major axis. The obliquity range is calculated using only the $|g|$ values for interactions of adjacent pairs of planets (e.g. TRAPPIST-1 c with TRAPPIST-1 b or with TRAPPIST-1 d) and the median obliquity for each range is plotted as a point on the range line. The planets with no possible obliquity for a given value of $|g|$ are plotted as a dashed line. The value for Earth's obliquity is also shown as a reference. There is a wide range of obliquity values possible for planets $0.01-0.4$ AU from their host star. Beyond roughly 0.4 AU, non-zero obliquity due to capture in CS2 is no longer excited.}
    \label{fig:obliquity_vs_a}
\end{figure*}

We have determined from the analyses described above that, similarly to studies of planets orbiting Sun-like stars, the relevant frequencies overlap at the $\sim50\%$ level when considered as a population (\citealt{Millholland19_obliquity}, see Figure \ref{fig:histogram_g_alpha}). We interpret this overlap as an indicator of the the common plausibility of the capture of planetary pairs into CS2. On a individual level, we consider each $\{\alpha,g\}$ pair separately, employing Equation \ref{eqn:obliquity} to estimate the resulting obliquity. In Figure \ref{fig:obliquity_vs_a}, we show our  predictions for obliquity from this analysis for all planets in the sample. We estimate a non-zero obliquity value for 216 of the 280 planets, or roughly $75\%$ of our sample. This higher percentage of predicted oblique planets (as opposed to the 50\% as an ensemble) is due to the favorable offset of $\alpha$ from $g$ shown in Figure \ref{fig:histogram_g_alpha}. While 50\% of the sample overlaps in frequency space (useful for consideration of commensurability at a population level), in fact $\alpha/|g| >1$ allows for obliquity excitation per Equation \ref{eqn:obliquity}, and this latter condition holds for 75\% of our sample. When combining the plausibility of excited obliquities together with low orbital eccentricity, it may be that tidal locking ought not necessarily to be the benchmark assumption for small, close-in planets orbiting M dwarfs.

\begin{figure*}[ht!]
    \centering
    \includegraphics[width=5in]{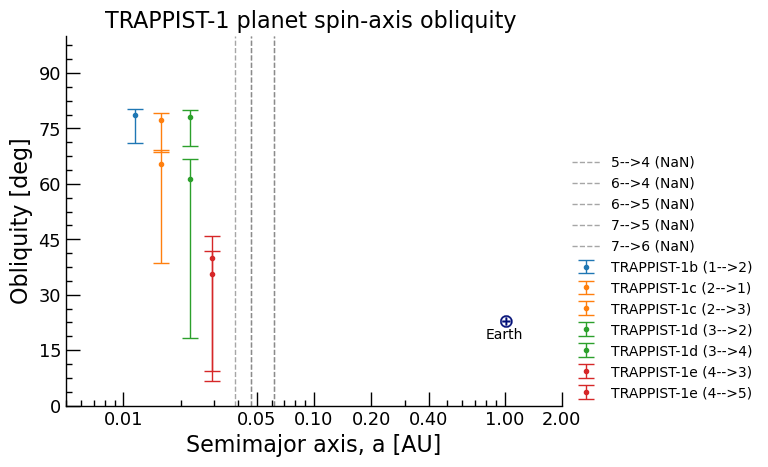}
    \caption{The four innermost planets in the TRAPPIST-1 system are amenable to capture into a non-zero obliquity state. We note that TRAPPIST-1 d and TRAPPIST-1 e (the green and red points, respectively) are in the habitable zone for the planet, which ranges from roughly $0.02-0.04$ AU.}
    \label{fig:trappistobliquity}
\end{figure*}

 We consider the individual case of TRAPPIST-1 \citep{Gillon16, Gillon2017, Luger2017, agol21} 
 to illuminate the relationship between the nominal ``habitable zone", and the range of orbital separations in which CS2 is plausible. In Figure \ref{fig:trappistobliquity}, using the steps outlined in Section \ref{sec:methods}, we show our obliquity predictions for its 7 planets. 
 Note that, given the definition of $g$ in Equation \ref{eq:g}, the predicted orbital precession depends on which pair of planets we consider. For planets with a neighbor on both sides, there result two predicted values for $g$. In the case of TRAPPIST, we consider both $g$ values for planets c, d, and e (TRAPPIST-b and TRAPPIST-h have only $g$, being on the respective edges of the system, and neither $g$ results in an excited obliquity for TRAPPIST-f and g). Both TRAPPIST-1 d and e have predicted non-zero obliquity and are in the plausible ``habitable zone" for their host star. 
 
 Given our order-of-magnitude approach, we have not considered here the details of the habitable zone \citep{Shields16}, but have simply identified ``habitability" to mean a fiducial 200-400 K equilibrium temperature (assuming an albedo of 0.3). We show all possible values for obliquity in Table \ref{tab:bigtable}. This result for TRAPPIST-1 highlights an important implication for M dwarf planet pairs trapped in CS2: it provides a potential mechanism for the maintenance of stable, long-term obliquity for planets \textit{in the habitable zone}. Around FGK dwarfs, in contrast, the orbital separations for which $\alpha/|g|$ supports plausible non-zero obliquity are much too close to the host star to plausibly be habitable.  

\begin{deluxetable*}{llllllllllll}
\tablecaption{Predicted values for planet $\alpha$, $g$, $\epsilon$, and $\omega_{\textrm{eq}}/n$ for the TRAPPIST-1 system. Values for $g$ are computed between pairs of planets in the system, $c \leftarrow b$, $c\rightarrow d$, $c \rightarrow e$ and so on. Because of this pairwise calculation, there are $n-1 = 7-1 = 6$ possible values for both $\epsilon$ and $\omega_{\textrm{eq}}/n$. \label{tab:bigtable}}
\tablecolumns{12}
\tabletypesize{\small}
\tablehead{ 
\colhead{Planet name} & 
\colhead{$\alpha_{\rm min}$ ($s^{-1}$)} & 
\colhead{$\alpha_{\rm max}$} & 
\colhead{g} & 
\colhead{$\epsilon_{\rm min}$} & 
\colhead{$\epsilon_{\rm max}$} & 
\colhead{$(\omega_{\textrm{eq}}/n)_{\rm min}$} & 
\colhead{$(\omega_{\textrm{eq}}/n)_{\rm max}$}
}
\startdata
TRAPPIST-1b & 2.4e-08 & 8.41e-08 & 4.64e-09 1 / s & 70.9 deg & 80.3 deg & 0.327 & 0.59  \\
   & '' & ''  & 3.95e-10 1 / s & 84.8 deg & 87.2 deg & 0.0968 & 0.18  \\
   & '' & ''  & 1.49e-10 1 / s & 86.8 deg & 88.3 deg & 0.0594 & 0.111  \\
   & '' & ''  & 6.7e-11 1 / s & 87.9 deg & 88.9 deg & 0.0399 & 0.0746  \\
   & '' & ''  & 4.02e-11 1 / s & 88.3 deg & 89.1 deg & 0.0309 & 0.0578  \\
   & '' & ''  & 7.35e-12 1 / s & 89.3 deg & 89.6 deg & 0.0132 & 0.0247  \\
TRAPPIST-1c & 6.02e-09 & 2.04e-08 & 4.64e-09 1 / s & 37.7 deg & 69 deg & 0.635 & 0.973  \\
   & ''  & '' & 1.42e-09 1 / s & 68.5 deg & 79 deg & 0.367 & 0.645  \\
   & ''  & '' & 3.51e-10 1 / s & 80 deg & 84.7 deg & 0.185 & 0.337  \\
   & ''  & '' & 1.33e-10 1 / s & 83.9 deg & 86.7 deg & 0.114 & 0.209  \\
   & ''  & '' & 7.43e-11 1 / s & 85.5 deg & 87.6 deg & 0.0853 & 0.157  \\
   & ''  & '' & 1.33e-11 1 / s & 88.1 deg & 89 deg & 0.0361 & 0.0665  \\
TRAPPIST-1d & 1.49e-09 & 5.25e-09 & 3.95e-10 1 / s & 67 deg & 78.6 deg & 0.38 & 0.678  \\
   & '' & '' & 1.42e-09 1 / s & 17.2 deg & 66.7 deg & 0.684 & 0.999  \\
   & '' & '' & 8.28e-10 1 / s & 51.6 deg & 73 deg & 0.539 & 0.896  \\
   & '' & '' & 2.46e-10 1 / s & 72.5 deg & 81.1 deg & 0.303 & 0.551  \\
   & '' & '' & 1.26e-10 1 / s & 77.9 deg & 83.7 deg & 0.217 & 0.402  \\
   & '' & '' & 1.49e-11 1 / s & 85.9 deg & 87.8 deg & 0.0753 & 0.141  \\
TRAPPIST-1e & 4.03e-10 & 1.42e-09 & 1.49e-10 1 / s & 61.6 deg & 76.4 deg & 0.446 & 0.775  \\
   & '' & '' & 3.51e-10 1 / s & 28.7 deg & 67.9 deg & 0.659 & 0.992  \\
   & '' & '' & 1.01e-09 1 / s & 6.04 deg & 41.8 deg & 0.958 & 1  \\
   & '' & '' & 5.73e-10 1 / s & 7.92 deg & 59.8 deg & 0.803 & 1  \\
   & '' & '' & 3.29e-10 1 / s & 33.9 deg & 68.8 deg & 0.64 & 0.983  \\
   & '' & '' & 3.92e-11 1 / s & 76.9 deg & 83.2 deg & 0.233 & 0.43  \\
TRAPPIST-1f & 1.1e-10 & 3.85e-10 & 6.7e-11 1 / s & 48.5 deg & 72 deg & 0.563 & 0.921  \\
   & '' & '' & 1.33e-10 1 / s & 19.3 deg & 62.9 deg & 0.755 & 0.998  \\
   & '' & '' & 2.46e-10 1 / s & 13.7 deg & 46.8 deg & 0.932 & 1  \\
   & '' & '' & 1.08e-09 1 / s & nan deg & nan deg & nan & nan  \\
   & '' & '' & 2.15e-09 1 / s & nan deg & nan deg & nan & nan  \\
   & '' & '' & 1.33e-10 1 / s & 18.3 deg & 62.8 deg & 0.757 & 0.999  \\
TRAPPIST-1g & 4.59e-11 & 1.51e-10 & 4.02e-11 1 / s & 27.9 deg & 66.9 deg & 0.679 & 0.992  \\
   & '' & '' & 7.43e-11 1 / s & 8.82 deg & 55.2 deg & 0.861 & 1  \\
   & '' & '' & 1.26e-10 1 / s & 6.2 deg & 32.5 deg & 0.986 & 1  \\
   & '' & '' & 3.45e-10 1 / s & nan deg & nan deg & nan & nan  \\
   & '' & '' & 2.15e-09 1 / s & nan deg & nan deg & nan & nan  \\
   & '' & '' & 1.7e-10 1 / s & nan deg & nan deg & nan & nan  \\
TRAPPIST-1h & 1.54e-11 & 5.38e-11 & 7.35e-12 1 / s & 56 deg & 74.3 deg & 0.505 & 0.852  \\
   & '' & '' & 1.33e-11 1 / s & 29.4 deg & 67.9 deg & 0.658 & 0.991  \\
   & '' & '' & 1.49e-11 1 / s & 14.3 deg & 66.3 deg & 0.691 & 1  \\
   & '' & '' & 3.92e-11 1 / s & 13.8 deg & 40.9 deg & 0.962 & 1  \\
   & '' & '' & 1.36e-10 1 / s & nan deg & nan deg & nan & nan  \\
   & '' & '' & 4.81e-10 1 / s & nan deg & nan deg & nan & nan  \\
\enddata
\end{deluxetable*}

\subsection{Inferred Planetary Rotation Rates}
\label{sec:rotation}

Planets in our sample with non-zero obliquity, provided they possess low eccentricity, experience subsynchronous rotation. 
Tidally-locked planets have a spin (rotation) rate which equals the mean-motion of the planet in its orbit, $\omega = n$, alternatively designated ``synchronous'' rotation. Earth, which has a spin rate much faster than its mean-motion ($15^\circ/hr$ and $0.04^\circ/hr$, respectively) would be ``super-synchronous'' and a planet with a spin rate slower than its mean-motion would be ``subsynchronous.'' It is this subsynchronous condition that we explore in the remainder of this paper. A planet captured into CS2 and excited to non-zero obliquity could experience subsynchronous rotation, where a ``day''  can last several planetary years. We calculate the subsynchronous rotation rate of the planet as a function of obliquity using Equation \ref{eq:epsilon_to_w}.

This is a compelling scenario to investigate because a planet orbiting very close to its M dwarf host would ordinarily rapidly become tidally locked and grind down to a state of permanent day and permanent night. However, if captured into CS2, such a planet could resist this effect for billions of years, maintaining some rotation with possibly significant implications for the planet's atmospheric circulation and heating. 

We have predicted that capture into CS2 can drive planets to high spin obliquities, and induce obliquity tides deep in the core of the planet. 

Using $n = 2\pi / P $ and $ \omega_{\textrm{eq}} = (\frac{\omega_{\textrm{eq}}}{n})(\frac{2\pi}{P}) $, we can write the formulae for $T_{\textrm{sidereal}}$ and $T_{\textrm{solar}}$ in terms of P, known for our sample, and $ \omega_{\textrm{eq}}/n$, which we computed previously.

\begin{equation}
    T_{\textrm{sidereal}} = \frac{2\pi}{\omega_{\textrm{eq}}} = \frac{2\pi}{(\frac{\omega_{\textrm{eq}}}{n})(\frac{2\pi}{P})} =  \frac{P}{\frac{\omega_{\textrm{eq}}}{n}}
    \label{eq:T_sidereal}
\end{equation}

\begin{equation}
    T_{\textrm{solar}} = \frac{T_{\textrm{sidereal}}}{1 - (T_{\textrm{sidereal}}/P)} 
    \label{eq:T_solar}
\end{equation}

\begin{figure}[ht!]
    \centering
    \includegraphics[width=3in]{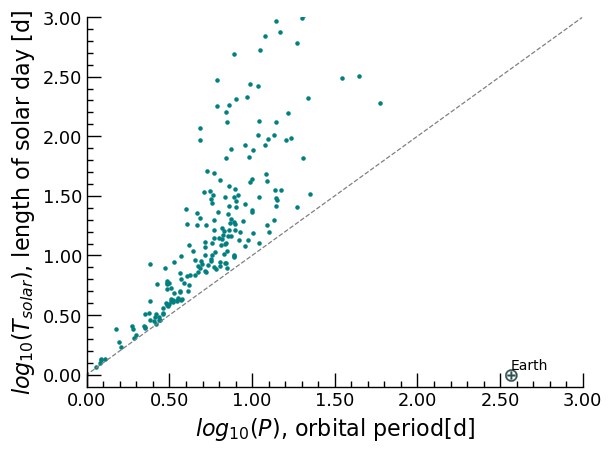}
    \caption{The length of a ``solar day'' lasts longer than a orbital year for planets captured into CS2. The horizontal axis shows the orbital period for the sample and the vertical axis shows the computed $T_{\textrm{solar}}$. The grey dashed line corresponds to $T_{\textrm{solar}}$ = P, or permanent day and permanent night for a tidally locked planet. Earth, shown for reference, has a $T_{\textrm{solar}}$ much smaller than its orbital period.}
    \label{fig:Tsol_P}
\end{figure}

We find that for planets in our sample which are captured into CS2 and thus not tidally locked, $T_{\textrm{solar}}$ corresponding to subsynchronous rotation is longer than the orbital period, P. The values for $T_{\textrm{solar}}$ are shown in Figure \ref{fig:Tsol_P}. For $P < 10$ days, $T_{\textrm{solar}}$ is roughly 1-2 times $P$, but can be much longer.

We can calculate the length of a sidereal day, $T_{\rm sidereal}$, how long it takes for the planet to complete one rotation, and use this and the orbital period, $P$, to calculate $T_{\rm solar}$, the solar day, which is how long it takes for a point on the surface of the planet to rotate to align with the host star again. On Earth, the solar day is only slightly longer than the sidereal day, but Mercury rotates subsynchronously and experiences a solar day much longer than its sidereal day. This effect for Mercury is not due to eccentricity tides, as with the planets in our M dwarf sample, but provides a useful analogue.

We compile all relevant calculated values for $\alpha$, the precession frequency; $g$, the nodal precession frequency;  $\epsilon$, the obliquity; and finally, the resulting $\omega_{\textrm{eq}}/n$, which is the ratio of equilibrium rotation rate $\omega_{\textrm{eq}}$ to mean-motion, $n$, for each planet. Table \ref{tab:bigtable} contains all resulting values for the TRAPPIST-1 system. 

\subsection{Long-term Stability of High-Obliquity Cassini State}
\label{ss:stable_cs2}

In this Section, we assess the likelihood of stability in the CS2 state to tidal dissipation using the formalism of \cite{Su2022a}. We have until this point considered only the spin and orbital precession frequencies, $\alpha$ and $g$, to assess plausibility of capture into the Cassini 2 state. However, many of the planets we consider orbit close enough to their M dwarf hosts to experience strong tidal forces which will drive the planetary spin rate towards synchronization and its obliquity towards zero. For this calculation, we adopt the use of $\eta=|g|/ \alpha$, the inverse of the quantity $\alpha/|g|$ considered above (in e.g. Figure \ref{fig:ag_vs_obliquity}), per its use in \cite{Su2022a}. For clarity, the condition for CS2 considered in the previous Sections ($\alpha/|g| >1$) corresponds to $\eta=|g|/\alpha<1$.

The persistence in CS2 depends on the strength of the obliquity tide: if sufficiently strong, it can suppress the excitation of obliquity. That is, at a given semi-major axis, the ``critical" tidal dissipation rate occurs when the tidal torque exceeds the maximum possible restoring torque (see also \citealt{Levrard07}). At this critical dissipation rate, CS2 becomes unstable. It must be compared to the characteristic tidal timescale $t_{\rm s}$ in Equation~\ref{eq:t_s}, the rate over which the spin orientation and rotational frequency of the planet evolve \citep{Su2022a}. If CS2 is stable, then the critical tidal dissipation timescale in Equation \ref{eq:t_s,c} must be \textit{longer} than the characteristic tidal timescale.

The characteristic tidal rate $t_{\rm s}$ is given by Equation \ref{eq:t_s} \citep{Su2022a}
\begin{equation}
    \frac{1}{t_s} \equiv \frac{1}{4C}\frac{3k_2}{Q}\left( \frac{M_*}{m_p} \right) \left( \frac{R_p}{a} \right)^3 n.
    \label{eq:t_s}
\end{equation}
We take $C = 0.4$ and $k_2/Q = 10^{-3}$, both of which are representative of planets in this mass range \citep[e.g.][]{Yoder95, lainey2016tidal}.

The critical tidal dissipation timescale $t_{\rm s,c}$ can be approximated as Equation \ref{eq:t_s,c} from \cite{Su2022a}
\begin{equation}
    t_{\rm s,c} \approx \frac{1}{|g| \sin{I}}\sqrt{\frac{2}{\eta_{\rm sync}\cos{I}}},
    \label{eq:t_s,c}
\end{equation}

where $I$ is the mutual inclination between the inner and outer planet. We consider two cases, $I= [2.5, 5.5]^{\circ}$, motivated by the typical range of mutual inclinations for \Kepler\ multi-planetary systems with a range of multiplicities \citep{ZhuDong2021, Fang12, Zhu18}. Typical M dwarf multiple planetary systems are characterized by a mutual inclination of 2$\pm$1$^{\circ}$ \citep{Ballard14}, closer to the lower end of the range we consider, though other studies of multiplanet systems have placed typical $I$ as high as 5$^{\circ}$ \citep{Figueira12}.

We go on to compare $t_{\rm s}$ to $t_{\rm s,c}$ for each of the planets in our sample. Quite conveniently, their ratio can be expressed simply:
\begin{align}
    \frac{1}{t_s}
        &=
            \frac{3}{2Q}\alpha_{\rm syn},\\
    \frac{t_s}{t_{\rm s, c}}
        &= 
            \eta_{\rm sync}^{3/2}
            \sin I
            \sqrt{\frac{\cos I}{2}}
            \frac{2Q}{3}.
\end{align}
Note that the requirement that $t_{\rm s} / t_{\rm s, c} > 1$, tidal stability of CS2, sets a lower limit on $\eta_{\rm sync}$, and, because obliquity is high for systems with $\eta_{\rm sync} < 1$, thus sets an upper limit on the maximum obliquity of planets in CS2.

\begin{figure}[tp]
    \centering
    {\begin{tabular}{cc}
         \includegraphics[width=0.45\linewidth]{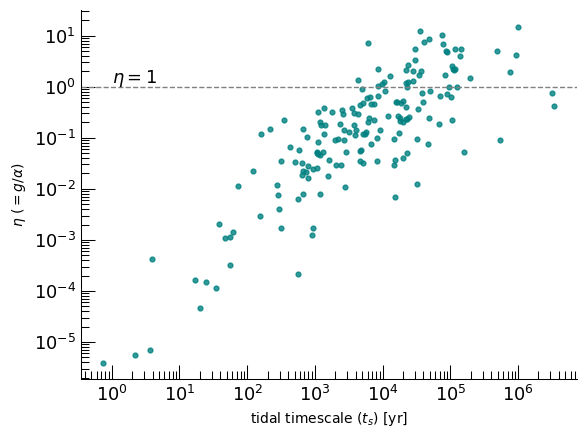} &
         \includegraphics[width=0.45\linewidth]{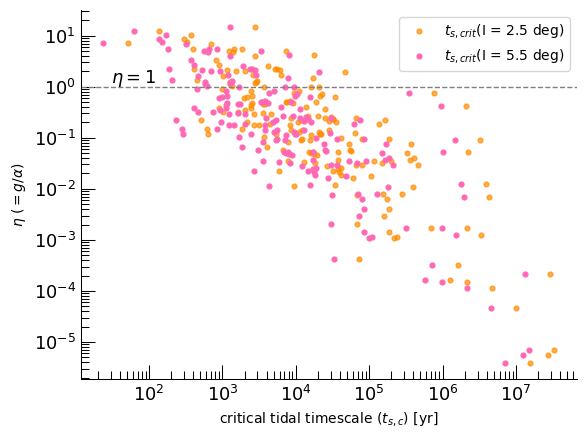}
    \end{tabular}}
    \caption{\textit{At left:} Characteristic tidal timescale $t_{\rm s}$ for all planets in sample, plotted against the ratio $\eta=|g|/\alpha$. We require $\eta < 1$ for excitation to a non-zero obliquity (see Equation \ref{eqn:obliquity}). We find that this condition is met more often for planets with shorter characteristic tidal timescales. \textit{At right:} Critical tidal timescale, $t_{\rm s,c}$, for all planets in sample, similarly versus $\eta$ (shown for two assumptions for mutual inclination $I$: 2.5$^{\circ}$ and 5.5$^{\circ}$). The opposite trend holds for planets that satisfy the $\eta < 1$ condition: they tend to have longer critical timescales.}
     \label{fig:timescales_eta}
\end{figure}

It is useful to consider how these two timescales typically cotrend with $\eta$, the ratio of $|g|/\alpha$, as they draw on many of the same variables. Figure \ref{fig:timescales_eta} depicts the relationship between each of these tidal timescales and $\eta$ (calculated for each planet per Section \ref{sec:estimation}).

\begin{figure}[ht!]
    \centering
    \includegraphics[width=4in]{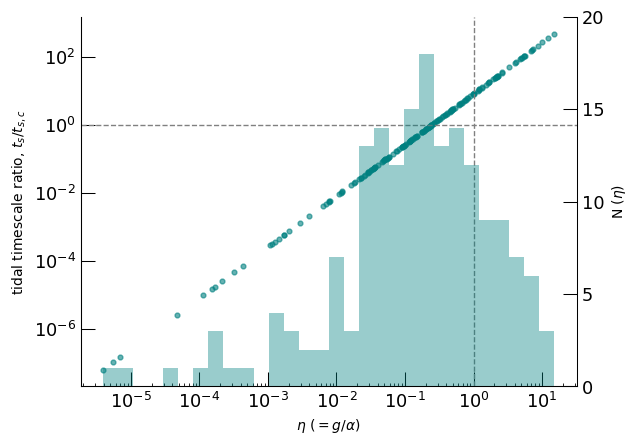}
    \caption{Ratio of characteristic tidal timescale to critical tidal timescale, $t_{\rm s} / t_{\rm s,c}$ (for $t_{\rm s,c}$ at I = 2.5$^{\circ}$), plotted against the ratio $\eta=|g|/\alpha$ for each planet we consider. The histogram of the $\eta$ values shows $\eta$ is most commonly between values of 0.01 and 1, where the points have the densest overlap. Above the horizontal grey dashed line, planets are tidally stable, with $t_{\rm s} / t_{\rm s,c}$. To the left of the vertical grey dashed line, $\eta < 1$. We can identify points in the upper left "quadrant" made by these intersecting dashed lines as the $36$ which satisfy both conditions for tidal stability and high obliquity.
    }
    \label{fig:tidal_ratio}
\end{figure}

As described above, CS2 is stable where $t_{\rm s} \geq t_{\rm s,c}$. Per Equation \ref{eq:t_s,c}, a higher mutual inclination (that is, a lower $\cos{I}$) will decrease the critical timescale, thus making it easier to satisfy the $t_{\rm s} \geq t_{\rm s,c}$ criterion necessary for stability. When we consider the higher mutual inclination $I$ of 5.5$^{\circ}$, we find that 93 of our sample of 280 planets could have stable CS2. In the less favorable case of $I$=2.5$^{\circ}$, we find that the number drops to 74 of 280 planets. In Figure \ref{fig:tidal_ratio}, we see that these planets are above the grey horizontal dashed line indicating $t_{\rm s} / t_{\rm s,c} =1$. When we filter this subset of planets to additionally be high-obliquity ($\eta < 1$, to the left of the vertical dashed line in Figure \ref{fig:tidal_ratio}), we find that 55 and 36 of the 280 in our sample satisfy both conditions for $I$ of 5.5$^{\circ}$ and 2.5$^{\circ}$, respectively. These findings are summarized in Table \ref{tab:filtertable} and Figure \ref{fig:tidal_ratio}.

\begin{deluxetable*}{ccc}
\tablecaption{Summary of applying tidal stability and high obliquity constraints to the sample of $280$ planets. \label{tab:filtertable}}
\tablecolumns{3}
\tabletypesize{\small}
\tablehead{ 
\colhead{I [deg] } & 
\colhead{($t_{\rm s} > t_{\rm s,c}$)} & 
\colhead{($t_{\rm s} > t_{\rm s,c}$) and ($\eta < 1$) }
}
\startdata
  $2.5^{\circ}$ & 74 & 36  \\
  $5.5^{\circ}$ & 93 & 55  \\
\enddata
\end{deluxetable*}

A consideration of Figure \ref{fig:timescales_eta} illustrates why the the $t_s \geq t_{\rm s,c}$ criterion is met so often: a high $t_s$ and a low $t_{\rm s,c}$ make the inequality easier to satisfy, and they tend to be associated with the same planets: those with higher values of $\eta$. Of course, once $\eta$ exceeds 1, no obliquity excitation is possible (see Equation \ref{eqn:obliquity}). For this reason, the dual conditions of favorable $\eta$, when combined with stability against tidal dissipation, tend to be associated with $\eta$ close to, but still less than, 1.

\subsection{Mechanisms for Capture into High-Obliquity Cassini State}\label{ss:capture_cs2}

Above, we have demonstrated that many of the planets in our sample have parameters compatible with a tidally-stable, high-obliquity CS2.
Next, we consider scenarios by which these planets can attain and retain their high obliquities to the present day. Similarly to Section \ref{ss:stable_cs2}, we employ $\eta=|g| / \alpha$ for the convention employed in \cite{Su2022a}. We consider two scenarios:
First, we consider the scenario of \citet{Millholland19_obliquity}, where the planet pair forms at wider separations, where $|g| / \alpha > 1$, and then subsequent disk migration brings the value of $|g| / \alpha$ to its smaller, present-day value.
Second, we consider the scenario of \citet{Su2022a}, where a planet with a large primordial obliquity can experience CS2 capture during the process of tidal realignment.



\subsubsection{Capture via Migration}

We begin by considering the \citet{Millholland19_obliquity} scenario. Recall first that the CS2 obliquity depends on the value of $|g| / \alpha$, which is given by the ratio of Eqs.~\eqref{eq:g} and~\eqref{eq:alpha_spin},

\begin{align}
    \frac{|g|}{\alpha} \simeq \frac{|g_{\rm LL}|}{\alpha}
        &= 
            \frac{b_{3/2}^{(1)}(\alpha_{12})\alpha_{12}^2}{2}
            \left(\frac{a_1}{R_p}\right)^3
            \frac{m_{1}}{M_\star}
            \frac{C}{k_2}
            \frac{n_1}{\omega}
            \left(\frac{m_2}{M_\star + m_1}
                + \frac{m_1}{M_\star + m_2}\frac{n_2}{n_1\alpha_{12}}\right)
            \nonumber\\
        &\approx
            \frac{3C}{2k_2}\left(\frac{a_1}{a_2}\right)^3
            \frac{m_1m_2}{M_\star^2}\left(\frac{a_1}{R_p}\right)^3
            \frac{n_1}{\omega}.\label{eq:g_alpha_diskmig}
\end{align}

First, since CS2 has a low obliquity and is tidally stable for $|g| / \alpha > 1$ \citep[e.g.][]{Ward1975a, Millholland19_obliquity}, a planet that forms in a system with $|g| / \alpha > 1$ is guaranteed to begin its evolution in CS2\footnote{
The critical ratio of $|g| / \alpha$ for guaranteed CS2 capture is $(\cos^{2/3}I + \sin^{2/3}I)^{-3/2} < 1$ \citep[e.g.][]{Peale1969, Ward1975a}.
Then, as the planetary system experiences orbital migration, the precession frequencies of the system change.
For simplicity, we assume that the planets undergo disk migration at the same migration rate, i.e.\ $\dot{a}_k = -a_k / \tau_a$ for some shared $\tau_a$.
As the $a_k$ decrease, Eq.~\eqref{eq:g_alpha_diskmig} shows that $|g| / \alpha$ decreases as well, and so the obliquity of CS2 increases. During this migration, the planet continues to librate about the CS2 equilibrium \citep[unless the planet reaches obliquities very near $90^\circ$,][]{2019millholland_disk, 2020sudisk}.
If feasible, this channel provides a guaranteed route to high obliquities for the inner planet.}

To assess the feasibility of this migration-driven obliquity excitation mechanism for the systems we consider, we evaluate the minimum amount of orbital migration required to realize these high obliquities.
To be precise, we denote $a_{\rm f}$ to be the present-day semi-major axis, and $a_0$ to be the \emph{smallest} initial semi-major axis (before orbital migration) such that the initial value of $|g| / \alpha$ (for a particular pair of planets) exceeds $1$.
Then, the ratio of $a_0 / a_{\rm f}$ quantifies the minimum amount of orbital migration that the planets in a system must have undergone such that the inner planet could have experienced guaranteed CS2 capture in the system's primordial architecture. We assume the ratio between the outer and inner planet semi-major axes remains constant between the start of migration and the present day: $a_2 / a_1 |_{\rm f} = a_2 / a_1 |_{0}$. Here, we are only considering planets with a present day $\eta < 1$ (CS2 is possible). Out of our sample of $280$ planets, $128$ meet this condition, with \textit{lower} values of $\eta$ corresponding to \textit{higher} values of obliquity.

\begin{figure}[tp]
    \centering
    {\begin{tabular}{cc}
        \includegraphics[width=0.4\linewidth]{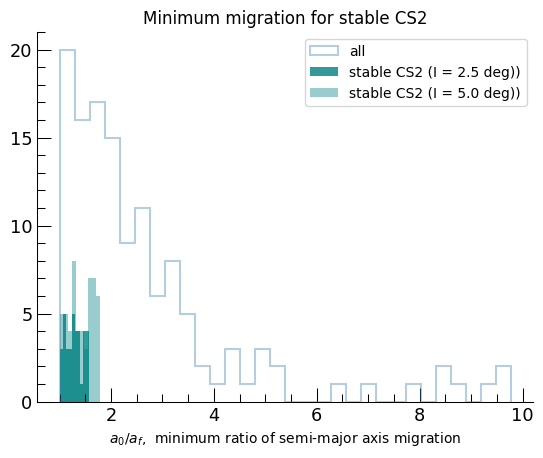}
        \includegraphics[width=0.45\linewidth]{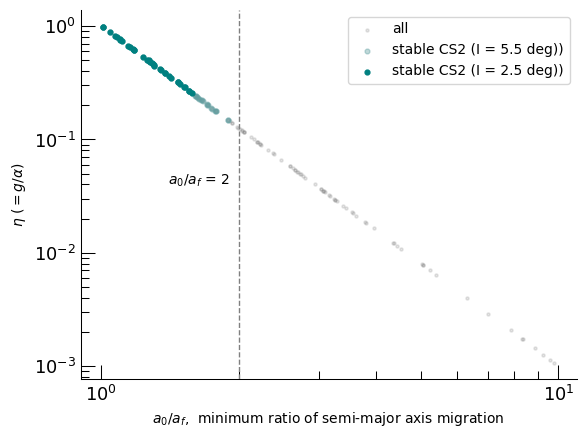}
    \end{tabular}}
    \caption{\textit{At left:} The grey histogram shows ratios of $a_0/a_f$, which we calculate for the 128 planets that already meet the $|g|/\alpha < 1$ condition in the present day. Lines are color-coded by the tidal stability condition calculated in Section \ref{ss:stable_cs2}, with grey corresponding to planets that do not meet the condition, teal corresponding to planets that meet the condition if mutual inclination $I=5.5^{\circ}$, and dark teal corresponding to planets that meet the condition if mutual inclination $I=2.5^{\circ}$. 
     \textit{At right:} For the same sample of planets, we depict the relationship between $\eta$ and $a_0/a_f$, with the same color-coding. The vertical grey dashed line shows $a_0/a_f = 2$. We conclude that planets likelier to be stable in CS2 require only modest amounts of migration.}    
     \label{fig:migration}
\end{figure}

The left panel of Figure ~\ref{fig:migration} illustrates the values of $a_0 / a_{\rm f}$ we obtain for our systems. Nearly all of the $a_0 / a_{\rm f}$ distribution falls below a factor of 4, indicating that typically modest migration is required, though there exists a longer tail out to $a_0 / a_{\rm f}\sim$10. The three histograms depict the planet samples relative to the tidal stability criterion from Section \ref{ss:capture_cs2}. For the tidally-stable planets, which are the most promising for the high-obliquity subsynchronous rotation scenario, we note that $a_0$ is not more than $\sim 2$ times the present-day $a$. This small $a_0/a$ ratio suggests that these planets do not require a dramatic migration history in order to be captured into a high-obliquity state. The right panel of Figure ~\ref{fig:migration} depicts how the sample of $a_0 / a_{\rm f}$ trends with $\eta$. We note that these tidally-stable planets span values of $\eta$ roughly between 0.1 and 1., bracketing a wide range of possible obliquities, close to $70^{\circ}$ for the smallest values of $\eta$ and less than $10^{\circ}$ for the values closest to 1. We conclude that capture into the tidally-stable CS2 does not require major inward migration of planets: most systems require orbital migration by less than a factor of $2$, especially those with present-day tidally-stable CS2.

\subsubsection{Capture from High Primordial Obliquity}

Next, we consider the tidal alignment route to CS2 capture, as described by \citet{Su2022a}.
In this scenario, the interior planet is assumed to experience a late stage of giant impacts and to form with a broad initial spin orientation. 
Then, for large initial obliquities, tidal dissipation causes the planetary spin to cross, and probabilistically be captured into, the high-obliquity CS2. While the dissipative resonance capture process for a given initial condition is complex, we can estimate the probability that a planet reaches the high-obliquity CS2 by averaging over initial spin orientations (obliquity and azimuthal phase) of the planet, assuming that it is isotropically distributed.
This permits us to estimate the probability that a given planet pair's inner planet is high-obliquity CS2 using the approximate expression \citep{Su2022a}:

\begin{equation}
    P_{\rm tCE2} \simeq \frac{4\sqrt{\eta_{\rm sync} \sin{I}}}{\pi} \left[ \sqrt{n/\omega_{0}} + \frac{3}{2 \left(1 + \sqrt{n/\omega_{0}} \right)}  \right].
    \label{eq:PtCE2}
\end{equation}

\begin{figure}[ht!]
    \centering
    \includegraphics[width=3in]{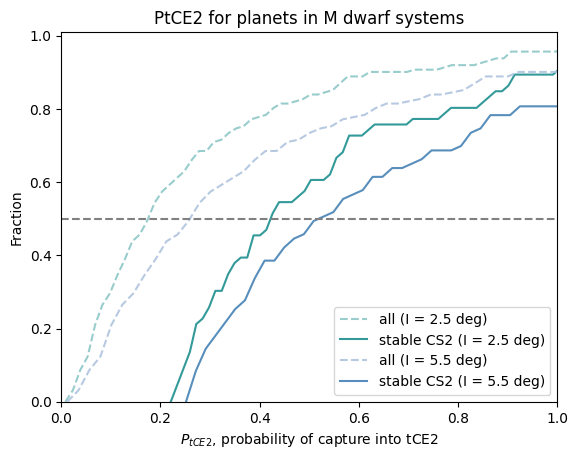}
    \caption{The cumulative distribution function (CDF) for the probability of evolution into tidal Cassini equilibrium 2 ($P_{\rm tCE2}$). Note that $P_{\rm tCE2}$ can exceed 1, at which point capture into CS2 is guaranteed (that is, the CDF jumps vertically to a value of 1.0 at $P_{\rm tCE2}=1.0$. The CDF is $\sim 20\%$ or higher for tidally-stable planets in our sample. We calculate $P_{\rm tCE2}$ for the top and bottom of our mutual inclination range, $I = 2.5$ and $5.5^\circ$, using an approximate analytical formula, Equation \ref{eq:PtCE2} from \citet{Su2022a} where $\eta_{\rm sync} \approx |g|/\alpha$; and $\Omega_{\rm s,i}$ is the planet spin frequency at the start of migration, which we assume to be $10n$, where $n$ is the mean-motion, $2\pi/P$. The dashed CDF curves are the $P_{\rm tCE2}$ values for all planets in the sample. The solid CDF curves are $P_{\rm tCE2}$ for tidally-stable systems, those where $t_{\rm s} \geq t_{\rm s,c}$. The horizontal grey dashed line is to guide the eye for $P_{\rm tCE2}$ values at the $50\%$ mark of each CDF. Values of $P_{\rm tCE2}$ can exceed $1$, meaning the planet is guaranteed to be captured into tCE2.}
    \label{fig:PtCE2}
\end{figure}

Here, $\omega_0$ is the initial spin rate of the planet, which we take to be $10n$ following \citet{Su2022a}. We note that it is possible for $P_{\rm tCE2}$ to exceed $1$, meaning the planet is guaranteed to be captured into tCE2. It is also the case that Eq.~\eqref{eq:PtCE2} is not accurate for large values of $\omega_0$, as it assumes that prograde obliquities damp, while the standard constant time lag tidal models result in substantial obliquity excitation for rapid initial spin rates \citep{Su2022a}. The inferred probabilities of the planets in our sample attaining high obliquities via this tidal realignment-driven resonance capture is shown in Fig.~\ref{fig:PtCE2}. Given that it is possible for $P_{\rm tCE2}$ to exceed a value of 1, the cumulative probability distributions shown in Figure ~\ref{fig:PtCE2} require brief explanation. They depict the cumulative probability corresponding to $0<P_{\rm tCE2}<1$. At the point of $P_{\rm tCE2}=1$ the CDF jumps vertically to 1.0, as any remaining planets have 100\% probability of capture. It can be seen in Figure~\ref{fig:PtCE2}, for example, that in half of all tidally stable systems (assuming mutual inclination $I=2.5^{\circ}$), capture probabilities exceed $\sim 50\%$. This CDF is equal to $\sim$0.75 at 1.0, at which point all remaining planets possess $P_{\rm tCE2}>1$.

The two capture mechanisms we have considered here are both subject to some uncertainties, which we discuss in Section~\ref{sec:conclusions}. 
Nevertheless, taken together, they suggest that a substantial fraction of the planet pairs in our sample can feasibly occupy the high-obliquity CS2 due to resonance capture in their past.

\subsection{Plausibility of Cassini State Capture in the Habitable Zone}
\label{sec:cassini_hz}

We predict roughly 
$20\%$, or 54 of 280,
of our sample will have $\epsilon > 10^{\circ}$ 
and be tidally stable, with $t_{\rm s} < t_{\rm s, c}$.

Of these 54 oblique, tidally-stable planets, 8-10 are within the planet equilibrium temperature range $T_{\rm eq} = 200-400$ K: comprising $20\%$ of this subset of the simplified fiducial ``habitable zone" and $3\%$ of our total sample. The stellar effective temperature range for this subset is $\teff = 2560-2860~K$, with a median temperature, $\teff = 3730~K$, lower than the median  for other stars in the sample, $\teff = 3910~K$. 


\begin{figure*}[ht!]
    \centering
    \includegraphics[width=5in]{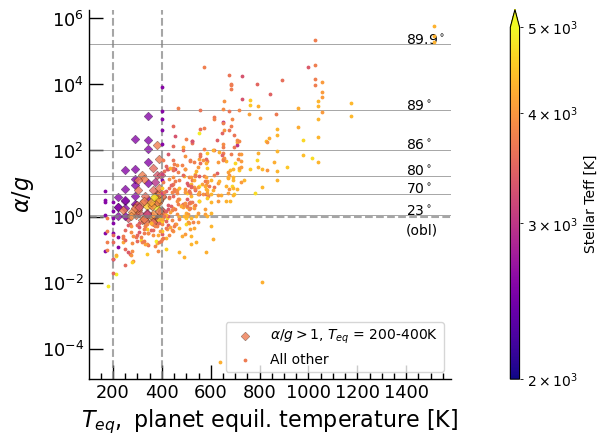}
    \caption{We predict that temperate planets are highly likely to be captured in a spin-orbit resonance that is amenable to exciting high obliquity. Further, we note that these temperate, oblique planets are primarily hosted by cool stars (median $\rm{T_{\rm eff}} = 3720~K$. The horizontal axis plots equilibrium temperature, $T_{\rm eq}$ in K and the vertical axis plots $\alpha/|g|$. Recall that values of $\alpha/|g| > 1$ permit obliquity excitation. The thin grey horizontal lines mark obliquity values corresponding to selected values of $\alpha/|g|$. Note $23^\circ$, Earth's obliquity, is shown. The color map corresponds to $\rm{T_{\rm eff}}$, the stellar effective temperature in K.  The dashed grey horizontal and vertical lines bound the region of $T_{\rm eq} = 200-400$ K and $\alpha/|g| > 1$.}
    \label{fig:aong_teq}
\end{figure*}

A summary of our findings at a stellar population level is shown in Figure \ref{fig:aong_teq}. Rather than depicting obliquity $\epsilon$ versus equilibrium temperature per se, we show $\alpha/|g|$ as a function of equilibrium temperature. In this framework, it's possible to see both the sample of planets plausibly excited to a high obliquity ($\alpha/|g|>1$) as well as those with no predicted excited obliquity ($\alpha/|g|<1$). For planets in the former sample, obliquity varies from $10-90^{\circ}$, which corresponds to subsynchronous rotation (as shown in Figure \ref{fig:weqn_obl}). The trend with stellar temperature is depicted here, as an approximate upper envelope to the function of $T_{\textrm{eq}}$ versus ($\alpha/|g|>1$). While stars with $T_{\textrm{eff}}>4000$ K tend to have $\alpha/|g|<1$ only for $T_{\textrm{eq}}>400$ K, at $T_{\textrm{eff}}\sim3000$ K we see temperature planets with high excited obliquities. For $T_{\textrm{eff}}\sim2000$ K, we predict nearly all planets with $200<~T_{\textrm{eq}}<400$ K to have $\alpha/|g|>1$ (indicating higher obliquity and thus subsynchronous rotation).

\section{Conclusions}
\label{sec:conclusions}

From an estimation of the relevant frequencies for known multi-planet systems orbiting M dwarf host stars, we have explored the plausibility of capture into CS2. The corresponding excitation of a non-zero obliquity over long timescales has important implications for our assumptions about the tidal locking of these planets. We find that commensurability of $\alpha$ and $g$ may be common, a necessary criterion for capture into CS2, though details rely on the evolution toward commensurability and the stability of the equilibrium state after capture \citep{Su2022a, Millholland19_obliquity}). We consider the problem at an order-of-magnitude scale: given our degree of uncertainty on individual planetary masses and tidal properties, we identify the following trends.

\begin{itemize}
    \item Roughly $50\%$ of the distributions for $\alpha$ and $g$ overlap in frequency space, as shown in Figure \ref{fig:histogram_g_alpha}). This overlap is more modest than the $80\%$ overlap calculated by \cite{Millholland19_obliquity} for FGK stars. This difference is due to an offset in $\alpha$ in our sample of planets toward higher values. The smaller overlap at a population level is useful for consideration of general commensurability, but the criterion for obliquity excitation ($\alpha/|g|>1$) is actually satisfied more often in our sample when planets are considered individually (75\%). 
    \item We identify moreover an overlap of the ``capture zone'' for CS2 (that is, the orbital separations where $g$ is most often near $\alpha$) and the habitable zone for planetary systems orbiting M dwarfs. Of our sample of 280 planets, 216 have obliquity values $\epsilon > 10^{\circ}$, and 46 of those 216 have a planet equilibrium temperature range $T_{\rm eq} = 200-400~K$. Of these 46, 8-10 are also tidally stable ($t_{\rm s} > t_{\rm s,c}$. 
    On the whole, the host stars for these systems have a median stellar effective temperature, $\teff = 3730~K$, lower than the median $\teff=3910~K$ for the other stars in the sample.  For stars with $T_{\textrm{eff}}\sim2000$ K, we predict nearly all planets with $200<T_{\textrm{eq}}<400$ K to have $\alpha/|g|>1$ (indicating higher obliquity and thus subsynchronous rotation). 

    \item Combining an assumption of low eccentricity together with our inferred obliquities for these planets, predict rotation rates to commonly be synchronous, as opposed to the common default assumption of 1:1 tidal locking. In degrees per hour (useful for benchmarking to Earth), we find $w_{\rm eq} = 0.025 - 3.92^\circ$/hr.
\end{itemize}

We conclude that capture into a high obliquity state, where subsynchronous rotation is possible, could in fact be a common phenomenon for temperate planets orbiting cool M dwarf stars.  Therefore, the widely-held assumption of tidal locking should be revisited in light of how common this spin-orbit resonant capture may be. Investigations along several axes can illuminate both the likelihood and the implications of our first-order findings reported here. 

We have explored the plausibility of capture into the high-obliquity Cassini State 2 (CS2) in terms of a simple overlap in frequencies $\alpha$ and $g$. In truth, not only whether but \textit{how} the spin and orbital precession frequencies cross each other as the system evolves in time is important.
Such convergence depends on the separate evolution of both frequencies.
We have explored two mechanisms by which these frequencies can be induced to converge.
One canonical picture involves $\alpha$ decreasing and $g$ increasing as the system ages: young, primordially spun-up planets start with high $\alpha$, while $g$ starts low due to the initially distant orbits. Disk migration of sub-Neptunes orbiting Sunlike stars \citep[e.g.][]{Millholland19_obliquity} illustrate how inward migration occurs and orbits shrink, so that typically $g$ increases, while $\alpha$ decreases as the planet spins down. The point at which the frequencies cross is where one or more of the planets may be excited to a high, but stable, obliquity.
A second picture involves keeping the planetary system's orbital architecture constant but assuming that the planets start with initially large obliquities.
Then, as the planets tidally spin down and realign, they cross the CS2 resonance \citep{Su2022a}, resulting in probabilistic resonance capture to the high-obliquity CS2 state.
Between these two mechanisms, we find that both can produce large-obliquity planets with quite modest assumptions: orbital migration by a factor of $\lesssim 2$ for the orbital migration scenario, and an isotropic initial spin distribution for the tidal realignment scenario.

We briefly discuss a few caveats to our obliquity capture scenarios above.
First, in evaluating the tidal stability of CS2 in Section~\ref{ss:stable_cs2}, we have followed existing literature and assumed that the tidal despinning and obliquity damping follow the standard constant time lag, equilibrium tidal friction models \citep{alexander1973, Hut81}.
With different tidal dissipation models that are more suitable for rocky planets, different spin-orbit resonances may appear \citep[e.g.][]{2012makarov_gj581d, 2022valente} and are suitable for further study.
Second, in evaluating the orbital migration-driven obliquity excitation scenario of \citet{Millholland19_obliquity}, we have assumed that the presence of the disk during the orbital migration process does not affect the CS2 capture process.
This simplification neglects effects such as mass accretion onto the planet and additional orbital precession due to the disk potential; preliminary work suggests that a more complete treatment of the disk does not affect the final resonance capture outcome significantly (Suar \& Millholland, in prep).
Third, in assessing the tidal realignment CS2 capture scenario of \citet{Su2022a}, we have assumed that the formation of the planetary systems in our sample result from a late stage of dynamical instability and giant impacts, which are thought to play a significant role in forming the observed super-Earth and sub-Neptune systems \citep[e.g.][]{2015inamdar, 2022goldberg_instab, 2023lammers_instrab}.
The resulting obliquities from such a phase of impacts has not been studied, though the isotropic distribution we adopt is likely representative of the true distribution.
We expect that these above caveats may change the quantitative number of high-obliquity planets in our sample, but not affect our qualitative conclusion that many planets in our sample are not currently tidally locked.


The resulting atmospheric circulation for planets captured into CS2 will be shaped from both above and below: the first critical effect of non-zero obliquity is the energy budget contributed from stellar photons incident on the top of the atmosphere (e.g. \citealt{Dobrovolskis13, Rauscher17, Wang16}. For an oblique planet (particularly if not tidally locked), the insolation pattern is complex: \cite{Ward74, Ward92} found that on planets with obliquities greater than the critical 54$^{\circ}$ (and less than 126$^{\circ}$), long-term average insolation at the poles is larger than insolation at the equator. The second effect of the the non-zero obliquity is the tidal heating to the planet’s interior \citep{Winn05_obliquity, Millholland19_obliquity}. This heating source in the planet’s interior contributes an energy flux from below that can reach a significant fraction of the insolation flux from above. Planets in high-obliquity states ought to experience heating up to several orders of magnitude higher than the zero-obliquity dissipation rate \citep{Millholland19_obliquity}. Indeed, this heating is sufficient to inflate the planetary radii to a detectable degree \citep{Millholland19_radius}.

Prospects for constraining exoplanetary obliquities in the coming years, using high-resolution spectroscopy \citep{Snellen14} and high-precision photometry to measure oblateness \citep{Seager02}, are likely limited to gas giant planets. The plausibility of residence in CS2 offers an intriguing possibility, even in absence of exact rotation rate information, that tidal locking may not be the default for habitable-zone M dwarf planets.


\section*{}
NMG is thankful for thoughtful discussions with Quadry Chance, Chris Lam, Sheila Sagear, and Billy Schap.
NMG and SAB acknowledge support for this work from the XRP program under NASA grant No. 80NSSC24K0152 and by the Heising-Simons Foundation through grant No. 2022-3533.

YS thanks Sarah Millholland for helpful and insightful discussions.
YS is supported by a Lyman Spitzer, Jr. Postdoctoral Fellowship at Princeton University.

We acknowledge that for thousands of years the area now comprising the state of Florida has been, and continues to be, home to many Native Nations. We further recognize that the main campus of the University of Florida is located on the ancestral territory of the Potano and of the Seminole peoples. 
As a Land Grant Institution, the University of Florida directly benefitted from the Land Grant College Act, otherwise known as the Morrill Act of 1862. Some of the `public' lands that were sold to fund the construction and expansion of the university were lands taken from the Seminole Nation as a result of the Seminole Wars. We, the authors, acknowledge the painful history of forced removal that occurred on these lands and the benefits that have been afforded to us as a result of the disenfranchisement, displacement, and death of so many. This history informs and shapes our work as astronomers who are committed to supporting all people under a common sky. 

This research has made use of the NASA Exoplanet Archive, which is operated by the California Institute of Technology, under contract with the National Aeronautics and Space Administration under the Exoplanet Exploration Program.


\facility {Exoplanet Archive}
\software{
    Astropy \citep{astropy:2013, astropy:2018},
    matplotlib \citep{matplotlib},
    Numpy \citep{numpy},
    pandas \citep{reback2020pandas, mckinney-proc-scipy-2010},
    pylaplace,
    Scipy \citep{2020SciPy-NMeth},
}

\bibliographystyle{aasjournal}
\bibliography{allrefs}

\end{document}